# SPACE—TIME—MATTER

Gerald E. Marsh

Argonne National Laboratory (Ret)

gemarsh@uchicago.edu

## ABSTRACT

This essay examines our fundamental conceptions of time, spacetime, the asymmetry of time, and the motion of a quantum mechanical particle. The concept of time has multiple meanings and these are often confused in the literature and must be distinguished if any light is to be thrown on this age-old issue. The asymmetry of time also has different meanings that depend on context—although the fundamental time asymmetry is associated with the expansion of the universe. These and related issues are discussed in both classical and quantum mechanical contexts.

**TABLE OF CONTENTS**

**Introduction**

The title of this essay is taken from Hermann Weyl's RAUM—ZEIT—MATERIE first published in 1921[1]. He began the book with some reflections of a philosophical nature: "Space and time are commonly regarded as the forms of existence of the real world, matter as its substance. A definite portion of matter occupies a definite part of space at a definite moment of time. It is in the composite idea of motion that these three fundamental conceptions enter into intimate relationship." Because the book was written at the beginning of the period when quantum mechanics revolutionized our conception of matter, the meaning of the second sentence in the quote has changed, but the last sentence in particular remains true today. How this is so is the subject of this essay.

**Spacetime**

It was the mathematician Hermann Minkowski who joined space and time together in his 1908 talk to the 80th Assembly of German Natural Scientists and Physicians [sic] stating that *"Henceforth, space by itself, and time by itself, are doomed to fade away into mere shadows, and only a kind of union of the two will preserve an independent reality."* Interestingly enough, Einstein was not initially comfortable with the reformulation of special relativity by Minkowski, his former teacher, into four-dimensional spacetime. Let us begin with this union of space and time.

In Euclidean space, which has a positive definite metric, the time coordinate has the same status as the space coordinates; in relativity theory, the time coordinate has a special status due to the indefinite metric of Einstein spacetime. The most important thing to remember is that, just like the space coordinates, the time coordinate itself is not associated with a "flow" in any particular time direction. It does not have an intrinsic orientation, asymmetry, or arrow associated with it. Put another way, there is no "arrow of time" associated with the time coordinate itself except for what we give it for illustrative purposes.

---

[1] H. Weyl, *SPACE—TIME—MATTER* (Dover Publications, Inc.)

The concept of "time" has multiple meanings: there is the coordinate itself; and there is the asymmetry of time in our three-dimensional space—which never changes its direction of flow; thermodynamic time, associated with the increase of entropy; psychological time, which each of us experiences as a present moment moving into the future; and finally, the concept of "cosmic time" associated with the expansion of the universe. Although these different concepts may be related, they are not identical and should not be confused. Nor should one confuse these different conceptions of time with the coordinate dependence of time measurements in relativity theory.

The Minkowski diagrams of special relativity are made up of a continuum of spacelike three-dimensional hypersurfaces along the time axis and perpendicular to it. The general view of time is that if one were to travel *backwards* in time one would see, for example, a sphere representing a propagating light pulse getting smaller and taking the size it had at an earlier time. That is, moving backward in time takes one to a three-dimensional space as it was in the past with the configuration of matter being what it was at each instant of past time. In this conception of time, three-dimensional hypersurfaces continue to exist in the sense that moving backward in time, were that possible, recapitulates three-dimensional space exactly as it was in the past. This concept of time leads to the usual conundrum that one could go back in time and murder one's grandfather. There is an even deeper problem.

The Einstein field equations of general relativity (the theory of gravity) have solutions that apply to objects like the earth or sun or to the universe as a whole. In the case of objects like the earth or the strong gravitational fields of massive neutron stars, these solutions have been tested to a very great accuracy. But the field equations also have perfectly good solutions, such as the infamous Gödel solution, that allow closed time curves.[2] Not only does this solution allow closed time curves, but in addition closed timelike curves pass through every point of this spacetime, and even more problematical is that there exists no embedded three-dimensional spaces without boundary that are spacelike everywhere, nor does a global Cauchy hypersurface exist.[3] Under the usual conception of time, moving in one direction along closed timelike curves is the equivalent to traveling backwards in time in that one may not only eventually arrive at the time when one began, but the configurations of three-dimensional space repeat themselves over and over again.

---

[2] This solution caused enormous ferment in physics and philosophical circles. See: P. Yourgrau, *Gödel Meets Einstein*: *Time Travel in the Gödel Universe* (Open Court, Chicago 1999); *A World Without Time: The Forgotten Legacy of Gödel and Einstein* (Basic Books, Cambridge MA 2005).

The Gödel solution and others like it are generally dismissed as being non-physical, but that simply begs the problem raised by their existence. The famous Kerr solution, representing the spacetime around a rotating mass, and which has no known interior solution—unlike the static Schwarzschild solution for a non-rotating mass—also has closed timelike curves if the angular momentum in appropriate units is greater than the mass, and one passes through the ring singularity. Yet, this solution is not dismissed as being non-physical.

Stephen Hawking has tried to get around the problem of closed time curves by introducing what he called the chronology protection conjecture: "*The laws of physics do not allow the appearance of closed timelike curves.*" But thus far there has been no proof of this conjecture. Einstein's field equations alone, being partial differential equations, only tell us about the value of a function and its derivatives in an arbitrarily small neighborhood of a point. Whether closed time curves exist or not is a global question that may also depend on the topology of the spacetime. Some things about closed time lines are known. For example, for asymptotically flat spacetimes, if certain energy conditions are satisfied, closed timelike curves can only occur if spacetime singularities are present.

If a signal may be sent between two points in spacetime only if the points can be joined by a non-spacelike curve, then the signal is said to be causal (this type of formulation allows for the possibility that the two points can only be joined by light rays). The spacetime will be causal if there are no closed non-spacelike curves. The non-rotating solutions to Einstein's field equations, such as the Schwarzschild and Friedman-Robertson-Walker cosmological solutions are causally simple. For most "physically realistic" solutions it has been shown that the chronology condition—that there are no closed timelike curves—is equivalent to the causality condition stating that there are no closed non-spacelike curves.[4]

---

[3] A global Cauchy surface is a spacelike hypersurface such that every non-spacelike curve intersects it only once.

More generally, the Einstein field equations belong to a class of partial differential equations known as symmetric hyperbolic systems.[5] Such equations have an initial-value formulation in the sense that once initial data are specified on a spacelike hypersurface, the subsequent time evolution follows from this data. Unlike the Gödel solution, where a global Cauchy hypersurface does not exist, if a Cauchy surface does exist, and initial conditions are imposed on it for its future evolution governed by the Einstein field equations, closed timelines—the equivalent of a "time machine"—cannot occur. As put by Geroch, ". . . there exist solutions of Einstein's equation in general relativity that manifest closed causal curves. But we do *not*, in light of this circumstance, allow observers to build time-machines at their pleasure. Instead, we permit observers to construct initial conditions—and then we require that they live with the consequences of those conditions. It turns out that a 'time-machine' is never a consequence, in this sense, of the equations of general relativity, . . ."[6]

The usual conception of time, with its past three-dimensional hypersurfaces that continue to exist, imposes our own psychological time. We can remember and think of how things were in the past, but this does not mean that the physical past continues to exist.

---

[4] While there is no need to discuss the time orientability of a spacetime here, it might be useful to give an example of a non-orientable spacetime. If one draws a circle representing the space axis along a Möbius strip and imposes a time direction perpendicular to the circle, after starting at any point and traversing the circle so as to return to the same point (on the other side of the strip of a paper model—a real Möbius strip only has one side), the time direction will be reversed. Such behavior implies that this 1+1 dimensional spacetime is not time orientable. While a spacetime that is non-orientable has a double covering space, which is orientable, that does not eliminate the problem in the underlying base space. Covering spaces are very useful in mathematics, but in terms of the physics, it does not seem to be possible to jump between the two spaces. Either the base space or the covering space must be chosen as representing physical spacetime.

[5] R. Geroch, "Partial Differential Equations of Physics", arXiv:gr-qc/9602055v1 (27 Feb 1996).

[6] R. Geroch, *Advances in Lorentzian Geometry: Proceedings of the Lorentzian Geometry Conference in Berlin*, M. Plaue, A. Rendall, and M. Scherfner, Eds. (American Mathematical Society 2011), p.59. arXive: gr-qc/1005.1614v1.

Another reason that the past, as conceived of in the usual conception of time, does not exist, and which will be discussed more thoroughly later in this essay in the subsection titled *Particles and their Motion*, has to do with the microreversibility—the symmetry under time reversal—of Schrödinger's equation for the wave function in quantum mechanics. For a system of particles this symmetry is generally broken; i.e., the equation of motion describes the history a system may follow, and this history will not in general coincide with the time-reversed evolution in typical quantum systems.

As pointed out earlier, just like spatial coordinates, the time coordinate itself is not associated with a "flow" in any particular time direction. If we measure the time distance around a closed timelike curve, there is no *prima facie* reason to expect the answer to be modulo the circumference.

Instead one may think of the evolution of time as being a one-dimensional covering space over the original closed timelike curve as shown in the Figure 1 below.[7] It is not necessary to identify the covering space as being the actual time curve in our universe since causality violations occur only if past three-dimensional hypersurfaces continue to exist. With this conception of time one could go around a closed time curve many times without a causality violation. The need for Hawking's chronology protection conjecture is eliminated.

---

[7] A relevant concept is called "unwrapping":S. Slobodov, "Unwrapping Closed Timelike Curves", *Found. Phys*. **38**, 1082 (2008). Unfortunately, the process of extending a spacetime containing closed timelike curves generally introduces other pathologies.

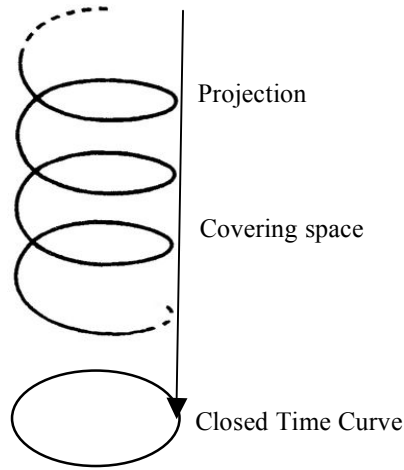

Figure 1. Closed time curve with a covering space. The original closed curve can be thought of as a projection of an *infinite* spiral over the closed time curve. Time changes monotonically in the covering space as one loops around the closed time curve.

**Time and the Expansion of the Universe**

The Friedmann-Lemaître spacetimes thought to represent our universe have exact spherical symmetry about every point, which implies that the spacetime is spatially homogeneous and isotropic admitting a six-parameter group of isometries whose orbits are space-like three-surfaces (constant time) of constant curvature (positive, negative, or flat). One may choose the coordinates such that the line element has the form $ds^2 = dt^2 - R^2(t)dl^2$, where $dl^2$ is the line element of a time-independent Riemannian three-space of constant curvature, be it positive, negative, or flat, and $R(t)$ is the expansion function. What this form of the metric tells us is that the proper physical distance $dl$ between a pair of commoving galaxies scales with time as $l(t) \propto R(t)$. For flat three-dimensional space, now believed to represent the actual universe, the function $R(t)$ monotonically increases with time.[8] One can readily show from the form of the metric that the velocity of separation of two commoving galaxies, $V$, is given by $V = [\dot{R}(t)/R(t)]l$, where the "dot" means the derivative with respect to time. This is the origin of the cosmological red shift. Thus, if $R(t)$ is constant, $V = 0$, and motion freezes.

The parameter $t$ of the Friedmann-Lemaître spacetimes is explicitly identified with the time parameter used to express physical relationships such as in Newton's and Maxwell's equations. This implies that if the time is set equal to a constant number so that the universe freezes at some radius, the time associated with physical processes also freezes—nothing can propagate or change in three-dimensional space. Motion and the flow of time are inexplicably linked, as pointed out by Weyl in the quote at the beginning of this essay. This would also be true in more general spacetimes where time may pass at different rates depending on the local mass-energy concentration. Thus, identification of the Friedmann-Lemaître time parameter (often called cosmic time) with physical time implies that the "flow" of time in three-dimensional space is due to the expansion of the universe.

---

[8] The flatness of three-dimensional space does not necessarily imply that the full spacetime is flat.

The connection would seem to be even deeper. These spacetimes begin with an initial singularity—the term being used loosely. This means space and time came into being together and, in the real as opposed to mathematical world, may not be able to exist independently. This is an obvious point: without time, there would be no space since expansion after the initial singularity would not be possible; and the expansion of space from the initial singularity implicitly introduces time and induces a time asymmetry in three-dimensional space.

This induced cosmic time is quite distinct from the thermodynamic time direction arising from increasing entropy when matter is present. It is also different from time as measured by "clocks" whose rate will vary according to both special and general relativity, but always in the implicit time direction induced in three-dimensional space by its expansion.

The Friedmann-Lemaître spacetimes can also have spatial sections that have positive curvature so that $R(t)$ is a cyclic function of time; i.e., the universe expands and then contracts (Fig. 2). But time is not reversed during the contraction phase; the initial asymmetry in time persists.[9] Thus, either the expansion or contraction of the universe leads to a time asymmetry in the same direction. The term "expansion" alone will continue to be used here since the real universe appears to be flat.

---

[9] In this connection I should mention the work of Hawking who considered quantum gravity and metrics that are compact and without boundary. He showed that the observed asymmetry of time defined by the direction of entropy increase is related to the cosmological arrow of time defined by the expansion of the universe. S.W. Hawking, "Arrow of Time in Cosmology", *Phys. Rev.* D **32**, 259 (1985).

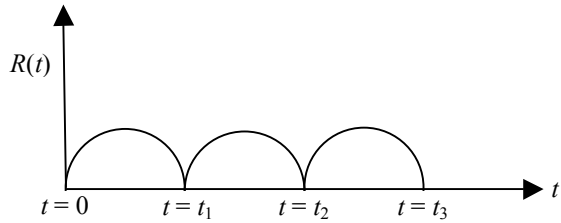

Fig. 2. The Cyclic Friedmann-Lemaître spacetime with positive curvature that first expands from an initial singularity and then contracts to a singularity. The equation for $R(t)$ is a cycloid.

If the singularities at the times indicated in Fig. 2 are identified so that one has a closed time curve and the time asymmetry persists in the same direction as indicated in the figure, then time could increase monotonically through the cycles (see Fig. 1 and associated discussion).

There is one additional point that should be made. The cosmological solutions to the Einstein field equations discussed above all have an initial singularity where spacetime itself is generally assumed to have come into being. The Einstein field equations themselves, however, do not inform us about what if anything existed "before" the initial singularity. The existence of the singularity simply indicates the limits of applicability of the field equations. In particular, these equations do not rule out the existence of some form of space or spacetime before the initial singularity. Most theoreticians assume that some form of quantum gravity will illuminate this issue. Unfortunately, current attempts in this direction—as exemplified by some forms of string theory, loop quantum gravity, non-commutative geometries, etc., have not had any convincing success. Also, to state a heretical view, there is no experimental evidence that space, time, or spacetime *is* quantized or that it need be quantized.[10] The desire to do so is primarily a matter of esthetics. It is based on the idea that because spacetime is a dynamical entity in its own right—due to its interactions with matter and energy—spacetime should in some sense be quantized.

---

[10] One often hears Heisenberg's uncertainty relations incorrectly raised in this context. But they have to do with the theory of measurement in quantum mechanics and are directly derivable from classical wave theory and the relations $E = h\nu$ and $p\lambda = h$, which will be discussed below. They do not apply to the fundamental limits of spacetime itself.

**The Asymmetry of Time**

The standard "big bang" model of cosmology assumes that at the very beginning of the universe there was no matter present but only energy in the form of enormously hot thermal radiation. The actual nature of this radiation, associated with a temperature of similar to $10^{32}$ °K at the Planck time of $10^{-24}$ sec, is not really known, although it is generally characterized as thermal radiation, which is, of course, of electromagnetic origin. The extremely hot origin of the universe is confirmed by the existence of the isotropic 3 °K background radiation.[11] The conversion of this early radiation into particle-antiparticle pairs, as the expanding universe cooled through a series of phase changes, is widely believed to be the source of the matter that exists today. The 3 °K background radiation itself comes from a time about half a million years after the initial singularity, by which time the plasma of ions (primarily hydrogen and helium, as well as electrons and photons) had formed and cooled to the point where it became a transparent gas. But there is a fundamental problem with this scenario that has not yet been resolved.

Consider the baryons (particles like neutrons and protons). From the observed ratio of the number of baryons to the number of photons in the background radiation—something like $10^{-9}$— it is apparent that only a small fraction of the matter survived the annihilation of the particle-antiparticle pairs. This means that somehow there must have been a small excess of matter over antimatter before the annihilation occurred. For this to be the case, the symmetry between baryons and antibaryons must be broken. Baryon number conservation must be violated so that the various allowed decay schemes resulting in baryons can lead to a difference between the number of baryons and anti-baryons. The criteria for breaking this symmetry was established by Sakharov[12] quite some time ago:

---

[11] See Appendix I.
[12] A. D. Sakharov, *Pisma Zh. Eksp. Teor. Fiz.* **5**, 32 (1967) [*JETP Lett.* **5**, 24 (1967)] [*Sov. Phys. Usp.* **34**,

both *C* and *CP* invariance must be violated, or otherwise for each process that generates a baryon-antibaryon asymmetry there would be a *C* or *CP* conjugate process that would eliminate the possibility of a net asymmetry; and there must be a departure from thermal equilibrium, or *CPT* invariance—which must hold for any local, relativistic field theory—would imply that there would be a balance between processes increasing and decreasing baryon number. There is some confusion in the literature about the meaning of the last requirement with regard to "time".

For example, Börner[13] states that, "Loosely speaking, the CPT-invariance of local, relativistic field theories and thermodynamic equilibrium imply the invariance under CP, because in thermodynamic equilibrium there is no arrow of time." Grotz and Klapdor[14] state that only if there is a departure from thermodynamic equilibrium will CP-violating interactions permit ". . . the rates of reactions which lead to the formation of baryons, to be larger than the rates of reactions which lead to antibaryons, but in thermodynamic equilibrium, no time direction is given, and the same would also apply to the inverse reactions."

Both statements argue that in thermodynamic equilibrium there is no Arrow of Time; i.e., no time direction is given. As it stands, this is certainly true, but as shown below in the discussion of thermodynamic time, this Arrow of Time has no relation to the kinematic time reversal transformation (see the Book by Sachs referenced below). There is often confusion between the Arrow of Time and *T*-violation. As put by Sean Carroll in the November 20, 2012 "blog" of the popular magazine *Discover*, referring to the recent results from BaBar on *T* violation, ". . . the entire phenomenon of *T* violation—**has absolutely nothing to do with that arrow of time** [emphasis in the original]."

---

392 (1991)] [*Usp. Fiz. Nauk* **161**, 61 (1991)]. Here *C*, *P*, and *T* are the discrete symmetries associated with charge, parity, and time respectively.

[13] G. Börner, *The Early Universe* (Springer-Verlag, Berlin 1993).

[14] K. Grotz and H.V. Klapdor, *The Weak Interaction in Nuclear, Particle and Astrophysics* (Adam Hilger, Bristol 1990).

With regard to Sakharov's requirement that there be a departure from thermodynamic equilibrium, Kolb and Turner[15] argue that, "The necessary non-equilibrium condition is provided by the expansion of the Universe. . . . if the expansion rate is faster than key particle interaction rates, departures from equilibrium can result." Calculations by Kolb and Turner show that only a very small *C* and *CP* violation can result in the necessary baryon-antibaryon asymmetry.

Systems in thermodynamic equilibrium while they do not have an Arrow of Time—called "thermodynamic time" in this essay in hopes of avoiding the kind of confusion found in the literature—do of course move through time in a direction given by the time asymmetry in the three-dimensional space within which we live.

Because of *CPT* conservation, it is clear that *CP* violation means that *T*-invariance is also violated. Now these symmetry violations are generally discussed in the context of particle decays. For example, the decay of the *K*-meson tells us that the violation of *T*-symmetry is very small. But no matter how small the breaking of time reversal invariance, the fact that it exists at all implies that there is a direction of time in particle physics; i.e., a time asymmetry, which—to reiterate it once again—has nothing to do with the thermodynamic Arrow of Time. Appendix 3 contains a discussion of CP violation and baryogenesis.

Before beginning the discussion of the asymmetry of time in quantum mechanics we turn to thermodynamic time so as to both complete the discussion above and introduce the Poincaré recurrence theorem.

**Thermodynamic Time**

Thermodynamic time has to do with the increase of entropy.[16] To begin with, the

---

[15]E. W. Kolb and M. S. Turner, *The Early Universe* (Addison-Wesley, New York 1990).

[16] An extensive and interesting discussion of time and entropy is contained in I. Prigogine, *From Being to Becoming: Time and Complexity in the Physical Sciences* (W. H. Freeman and Company, 1980).

Poincaré recurrence theorem,[17] associated with thermodynamic and classical systems in general, states that for an isolated and bounded non-dissipative system any particular state will be revisited arbitrarily closely; for macroscopic systems composed of many particles the recurrence time will be very, very large. A simple example is a perfect gas confined to one side of a chamber by a membrane with the other side of the chamber being evacuated. If a hole in the membrane is opened, the gas will flow into the vacuum side; but ultimately all the gas will return to its original configuration after the Poincaré recurrence time has elapsed. From the point of view of thermodynamic time, it is possible to return to where the physical configuration of matter is *arbitrarily close* to its original configuration provided the assumptions given above on the nature of the system hold. What has been called cosmic time above always increases monotonically into the future even for such systems.

Let us explore this issue more quantitatively. Consider a one dimensional lattice of $N$ particles of mass $m$ elastically coupled to their nearest neighbors by springs with a force constant $K$, and let one particle have a mass $M \gg m$, which at $t = 0$ is given some velocity, the other particles being at rest. Rubin[18] computed the subsequent motion of the lattice and for large $N$ found that the motion of the single particle with mass $M$ was damped nearly exponentially. But the time symmetry is preserved and after a time $\frac{N}{2}\left(\frac{m}{K}\right)^{1/2}$ the lattice system completes a Poincaré cycle and returns to the original configuration at $t = 0$. A similar effect occurs with quantum systems as will be shown later in this essay.

While a bounded system may therefore return to its initial state, there is no asymmetry in time involved. Nonetheless, one often hears of the "thermodynamic arrow of time" established by the second law of thermodynamics and the increase of entropy. The situation with thermodynamic time is quite murky. As put by Brown and Uffink,[19] "All

---

traditional formulations of the Second Law presuppose the distinction between past and future (or 'earlier' and 'later', or 'initial' and 'final'). To which pre-thermodynamic arrow(s) of time were the founding fathers of thermodynamics implicitly referring? It is not clear whether this was a question they asked themselves, or whether, if pushed, they would not have fallen back on psychological time."

That idea that the thermodynamic arrow of time coincides with the psychological arrow of time led Hawking to observe that ". . . the second law of thermodynamics is really a tautology. Entropy increases with time, because we define the direction of time to be that in which entropy increases." [20] There has been some objection to this pithy characterization of the second law, but it suffices for the purposes of this essay. The connection between the thermodynamic arrow of time and the physics of time reversal has been put quite succinctly by Sachs, ". . . the Arrow of Time has little to do with the time variable as measured by physicists. In particular it has no bearing on the physics of time reversal." [21] The thermodynamic arrow of time will play no further role in this essay. On the other hand, the Poincaré recurrence theorem will appear again in the next section.

**Time Asymmetry in Quantum Mechanics**

Below, in discussing Feynman's picture of the scattering of the Dirac wave function by a potential, waves will be allowed to travel backwards in time. These waves correspond to the negative energy states of the Dirac equation. That is, positrons may be interpreted as electrons propagating backwards in time. This may be explicitly shown by the transformation properties of the Dirac equation under the combination of parity, charge conjugation, and time reversal.

The Feynman interpretation of a positron as a backward in time moving electron is not inconsistent with the interpretation of time given above where past three-dimensional spacelike hypersurfaces do not continue to exist. The propagation into the past is very limited and the Feynman interpretation only applies to elementary particles. One way to accommodate this is to think of the three-dimensional space or hypersurface within which we live as having a very small thickness in the time dimension.[22]

---

[20] S. Hawking, *The No Boundary Condition And The Arrow Of Time*, in J. J. Halliwell, J. Pérez-Mercador, and W. H. Zurek, (eds), *Physical Origins of Time Asymmetry* (Cambridge University Press, Cambridge 1994).

[21] R. G. Sachs, *op. cit.*

To simplify the discussion of time asymmetry in quantum mechanics, let us consider the Schrödinger equation $H|\Psi\rangle = i\partial_t|\Psi\rangle$. Like the Dirac equation the probability amplitude $\Psi$ is invariant under the $T$ operator so that the physical content of the theory is unchanged. What will be shown is that even though the physical content of quantum mechanics is preserved under time reversal (micro-reversibility under the $T$ operator), when one considers multiple systems an asymmetry in time results. The discussion here follows that given by Davies.[23]

The Poincaré recurrence theorem associated with thermodynamic and classical systems that was discussed above has a quantum mechanical analog: Consider a collection of systems having only a ground state and one excited state whose energy can vary with the system. Now assume all systems are in their ground states save for one that is in its excited state. Assume further that all the systems are coupled by an interaction Hamiltonian $H_{int}$. After some time passes there is a probability that the original excited system is in its ground state and one of the other systems is in its excited state. Davies finds that for two identical coupled systems, the Schrödinger equation gives a probability amplitude for the original excited system of $\cos^2(|H_{int}| t)$. Here the Poincaré recurrence period is $2\pi/|H_{int}|$. On the other hand, for a large number of systems the probability amplitude for the original excited system is $e^{-2\pi|H_{int}|^2 t/\Delta E}$, where $\Delta E^{-1}$ is the density of states available. This is the usual time asymmetrical decay of an excited state with a half-life of $\Delta E/(2\pi|H_{int}|^2)$. As the number of systems increases, the density of states available goes to zero and the probability of the original state returning to its original excited state tends to zero.

---

[22] R.P. Feynman, *Quantum Electrodynamics* (W.A. Benjamin, Inc., New York 1962), pp.84-85.
[23] P. C. W. Davies, The Physics of Time Asymmetry (University of California Press, Berkeley 1977), § 6.1.

While quantum mechanics satisfies what is known as the principle of micro-reversibility, processes that appear asymmetric in time are related to special initial conditions and the openness of the system, a good example being radioactive decay.

**Particles and their Motion[24]**

"Matter", the third element of Weyl's book title, and its motion in the spacetime continuum, requires a quantum mechanical approach as opposed to using the conceptual particles of classical physics. The reason is that there is no classical world—only a many-particle quantum mechanical one that, because of localizations[25] due to interactions, allows the emergence of the classical world of human perception. The quantum mechanical approach also allows a more fundamental understanding of the concept of motion, one that transcends that of classical mechanics. The Standard Model of particle physics deals with the basic nature of matter. Here, matter will be discussed only with regard to the motion of a quantum mechanical particle such as an electron or proton.

---

[24] This section contains a modified and expanded portion of material taken from G.E. Marsh, "Quantum Mechanics and Motion: A Modern Perspective", *Physics Essays* **23**, 2 (2010). That paper's primary objective was to explore the question of how forces affect the probability distribution of a quantum mechanical particle and how the motion of macroscopic objects is governed by the quantum mechanics of its constituent particles and their interactions with each other.

[25] The use of the term "localization" is deliberate. There is no need to bring in the concept of measurements with its implicit assumption of the existence of an "observer." It is not necessary that the fact that an interaction has occurred somehow enter human consciousness in order for the particle to be localized in space and time. The argument that it must enter human consciousness has been used, for example, by Kemble [E. C. Kemble, *The Fundamental Principles of Quantum Mechanics with Elementary Applications*, Dover Publications, Inc., 1958, p. 331] who states that "If the packet is to be reduced, the interaction must have produced knowledge in the brain of the observer. If the observer forgets the result of his observation, or loses his notebook, the packet is not reduced." It is not our purpose here to enter into a discussion of quantum measurement theory, but interpretations such as that expressed by Kemble often—but not always—rest on a lack of clarity as to what the wave function is assumed to represent. That is, whether the wave function applies to a single system or only to an ensemble of systems. While the ensemble interpretation has proven conceptually quite valuable in a number of expositions of measurement theory, it is difficult to understand how the wave function cannot apply to an individual system given the existence of many interference experiments using a series of *individual* electrons—where each electron participating in the production of the interference pattern must interfere with itself. Perhaps the most well known attempt to bring consciousness into quantum mechanics is that of Eugene Wigner. The interested reader is referred to Wigner's book *Symmetries and Reflections* (Indiana University Press, Bloomington & London 1967), Section III and references therein.

Earlier it was stated that in relativity theory, the time coordinate has a special status due to the indefinite metric of Einstein spacetime. This indefinite metric also results in the speed of light being the maximum velocity of propagation.[26] If a particle or wave were to move faster than the speed of light, there would be a reference frame within which the particle would appear to be moving into the past. By considering a charged particle, Feynman, in his essay *The Reason for Antiparticles*,[27] used this property of an indefinite metric to show that antiparticles must exist. He did this by requiring that only positive energy be allowed in the quantum mechanical wave function associated with a particle and used the fact that the Fourier transform of a function restricted to only positive frequencies (positive energy) cannot vanish anywhere; while a wave function composed of only positive frequencies nowhere vanishes outside the light cone, it does become small over a distance comparable to the Compton wavelength. Those who remember the calculation of the Fourier decomposition of a "top-hat function", which is non-zero over only a finite range, might find this surprising, but the Fourier representation of such a function involves both positive and negative frequencies. Because the wave function of a particle composed of only positive frequencies is non-zero outside the light cone, there is a reference frame where the particle will appear to be moving backward in time. To summarize, the restriction to positive energies and the requirements of special relativity result in an interpretation where particles can move backward in time. An identity is made between the negative energy components of the scattered wave and the waves traveling backward in time.

The identification of particles moving backward in time as antiparticles—an interpretation discussed by Stückelberg, Feynman, and others—is fully compatible with the concept of the "flow" of time being the result of the expansion of the universe as introduced above, and—as discussed earlier—it is consistent with the non-existence of past three-dimensionsl hypersurfaces.  Let us turn now to the foundations of quantum mechanics.

---

[26] Newton's mechanics and Coulomb's law of attraction or repulsion between electric charges rely on *instantaneous* action at a distance. These laws were found to be in disagreement with phenomena associated with rapidly moving objects or charges. The cure was to introduce the concept of a field; and because the gravitational and electromagnetic fields carry energy, there must be a maximum velocity of propagation if the energy is to be finite. If $l$ is the distance traveled, and $c$ the velocity of propagation, we have $\Delta l = c\Delta t$, or $\Delta l^2 - c^2\Delta t^2 = 0$, which implies that one has an indefinite metric. This is generalized to $\Delta s^2 = \Delta l^2 - c^2\Delta t^2$ and $\Delta s^2 = 0$ corresponds to the maximum velocity of propagation. See the introductory parts of: A. Einstein, *The meaning of Relativity* (Princeton University Press, Princeton 1950.

[27] R. P. Feynman and S. Weinberg, "Elementary Particles and the Laws of Physics: The 1986 Dirac Memorial Lectures" (Cambridge University Press, Cambridge 1993).

The origination of quantum mechanics dates back to Max Planck in 1900 and his studies of heat radiation that led him to introduce the postulate that energy came in discrete, finite quanta of energy $h\nu$.  Planck was awarded the 1918 Nobel Prize in Physics for his work but was never comfortable with the idea of quanta.  Nonetheless, essentially all of quantum theory follows from special relativity and Planck's discovery that $E = h\nu$.

It is often forgotten, however, that in 1906 Einstein pointed out that Planck's reasoning in reaching his famous formula was inconsistent. This did not mean, however, that Planck's "quantum theory" had to be rejected.  Using the conservation of energy and the assumption that the entropy of cavity radiation was at a maximum, Einstein was able to show that radiation was composed of a finite number of localized energy quanta $h\nu$.[28] Since photon-photon interactions are essentially negligible, the derivation also implicitly assumes that matter is present in the cavity since radiative equilibrium could not be reached otherwise.  So what the derivation actually shows is that if one has radiative equilibrium $E = nh\nu$.

It was de Broglie, in his 1924 publication "*Recherches sur la Théorie des Quanta*", that introduced the thesis that elementary particles had associated with them a wave, what we call the wave function, and what de Broglie called an "*onde de phase*" or a *phase wave*. It is a consequence of the relation $E = h\nu$.  In his 1929 Nobel lecture he used the following argument:

$$p = \gamma m v = \gamma m c^2 \frac{v}{c^2} = E \frac{v}{c^2}.$$

He now identifies the energy $E$ of a *massive* particle with $E = h\nu$ to give

$$p = \frac{h\nu}{c^2/v}.$$

---

[28] There is a very good discussion of the derivation and history in Max Jammer, *The Conceptual Development of Quantum Mechanics* ( McGraw-Hill, New York 1966).

This identification is the key step used by de Broglie in deriving his relation. Since the velocity of the massive particle is always less than that of light, so that $c^2/v > c$, he states that "*qu'il ne saurait être question d'une onde transportant de l'énenergie*" (it cannot be a question of a wave transporting energy). Consequently, he makes another key assumption that $c^2/v$ corresponds to a phase velocity via $v v_{ph} = c^2$, so that

$$p = \frac{h}{v_{ph}/v}.$$

Since $v_{ph} = \nu\lambda$, de Broglie obtains his fundamental relation $\lambda p = h$. Notice that for $p = 0$, the wavelength is infinite, which implies that there is no oscillation and thus no phase wave. What this tells us is that de Broglie's phase wave is related to a particle's motion through space and time. However, the lack of oscillation for $p = 0$ is a result of the exclusion of relativistic effects. A relativistic formulation would show that when a particle is stationary, it has a frequency of oscillation associated with it called the *zitterbewegung*, which de Broglie thought of as the inherent frequency of the electron.

In more modern terms, a spin ½ massive particle like the electron can be viewed as oscillating between a left-handed massless particle with helicity +½ and a right-handed massless particle with helicity −½. Each of these is the source for the other with the coupling constant being the rest mass.

The two relations $E = h\nu$ and $\lambda p = h$ allow the derivation of the relations $E = i\hbar \frac{\partial}{\partial t}$ and $\boldsymbol{p} = -i\hbar \nabla$ used in the Dirac or Schrödinger equations, so it should be no surprise that the wave functions (*ondes de phase*) they determine are not restricted to the interior of the light cone. This becomes especially evident when considering the decoherence of correlated (entangled) quantum systems.

For the purposes of this essay, massive particles will be defined as excitations of spacetime having the usual properties of quantum mechanical particles that may be localized by interactions with each other or by interaction with fields. The situation is more complicated for massless quantized fields; the photon, for example, is not localizable.[29]

The concept of localization by interactions is completely consistent with the collapse of the wave function as found in the Copenhagen interpretation of quantum mechanics and should not be confused with what has come to be known as the "decoherence program". The latter assumes that all quantum systems are entangled with their environment and that a quantum mechanical particle becomes classical by environmentally induced decoherence. "However, since only unitary time evolution is employed, global phase coherence is not actually destroyed—it becomes absent from the local density matrix that describes the system alone, but remains fully present in the total system-environment composition. . . . The selection of preferred sets of states . . . are determined by the form of the interaction between the system and its environment and are suggested to correspond to the "classical" states of our experience."[30] The decoherence approach to quantum mechanics uses a type of "coarse-graining" to average over environmental states whose details are not of importance. The result is a "reduced density matrix" obtained by averaging over environmental states. This is very different from the localization via interactions discussed here.

To deal with the concept of motion we must begin with the well-known problem of the inconsistency inherent in the melding of quantum mechanics and special relativity. One of the principal examples that can illustrate this incompatibility is the Minkowski diagram, where well-defined world lines are used to represent the paths of elementary particles while quantum mechanics disallows the existence of any such well-defined world lines.

---

[29] This has been discussed extensively by Henri Bacry [*Localizability and Space in Quantum Physics*, Lecture Notes in Physics, Vol. 308, Springer-Verlag 1988].

[30] Maximilian Schlosshauer, "Decoherence, the measurement problem, and interpretations of quantum mechanics" *Rev. Mod. Phys.* **76**, (October 2004). The effect of environmental decoherence has also been discussed by Zurek and Halliwell: W. H. Zurek, "Decoherence and the Transition from Quantum to Classical", *Physics Today* (October 1991); J. J. Halliwell, "How the quantum universe became classical", *Contemporary Physics* **46**, 93 (March-April 2005)..

Feynman[31] in his famous paper "The Theory of Positrons" partially avoids the above conundrum, implicit in drawing spacetime diagrams, by observing that solutions to the Schrödinger and Dirac equations can be visualized as describing the scattering of a plane wave by a potential. In the case of the Dirac equation, the scattered waves may travel both forward and backward in time and may suffer further scattering by the same or other potentials. While one generally does not indicate the waves, and instead draws world lines in Minkowski space between such scatterings, it is generally understood that the particle represented by these waves does not have a well-defined location in space or time between scatterings.[32]

The Feynman approach visualizes a non-localized plane wave impinging on a region of spacetime containing a potential, and the particle the wave represents being localized to a finite region of Minkowski space by interaction with the potential. The waves representing the scattered particle subsequently spread through space and time until there is another interaction in the same potential region or in a different region also containing a potential, again localizing the particle. Even this picture is problematic since the waves are not observable between interactions.

For the Dirac equation, the now iconic Figure 3 is intended to represent electron scattering from two different regions containing a scattering potential. The plane electron wave comes in from the lower left of the figure, is scattered by the potential at A(3). (a) shows the scattered wave going both forward and backward in time; (b) and (c) show two second order processes where (b) shows a normal scattering forward in time and (c) the possibility of pair production.

---

[31] R. P. Feynman, *Phys. Rev.* **76**, 749 (1949).

[32] This is best exemplified by the path integral formulation of non-relativistic quantum mechanics. The latter also has the virtue of explicitly displaying the non-local character of quantum mechanics. The history has been well summarized by Richard MacKenzie (arXiv:quant-ph/0004090 v1 24 Apr 2000): "In 1933, Dirac made the observation that the action plays a central role in classical mechanics (he considered the Lagrangian formulation of classical mechanics to be more fundamental than the Hamiltonian one), but that it seemed to have no important role in quantum mechanics as it was known at the time. He speculated on how this situation might be rectified, and he arrived at the conclusion that (in more modern language) the propagator in quantum mechanics 'corresponds to' $\exp iS/\hbar$, where $S$ is the classical action evaluated along the classical path. In 1948, Feynman developed Dirac's suggestion, and succeeded in deriving a third formulation of quantum mechanics, based on the fact that the propagator can be written as a sum over all possible paths (not just the classical one) between the initial and final points. Each path contributes $\exp iS/\hbar$ to the propagator. So while Dirac considered only the classical path, Feynman showed that all paths contribute: in a sense, the quantum particle takes all paths, and the amplitudes for each path add according to the usual quantum mechanical rule for combining amplitudes. Feynman's original paper, which essentially laid the foundation of the subject . . . was rejected by Physical Review!"

Feynman meant this figure to apply to a virtual process, but—as discussed by Feynman—with the appropriate interpretation it applies to real pair production as well. Although the lines are drawn to represent these particles, no well-defined world lines exist.

Since waves scattered into the past and future represent the *probability* of where a particle could be found after a series of interactions, it is not possible to have a time-reversed evolution that would exactly recapitulate the interaction history of the particle. This is consistent with the concept of time discussed earlier in this essay where it is not possible to travel backwards in time to a previous three-dimensional hypersurface where the configuration of matter would be as it was in the past.

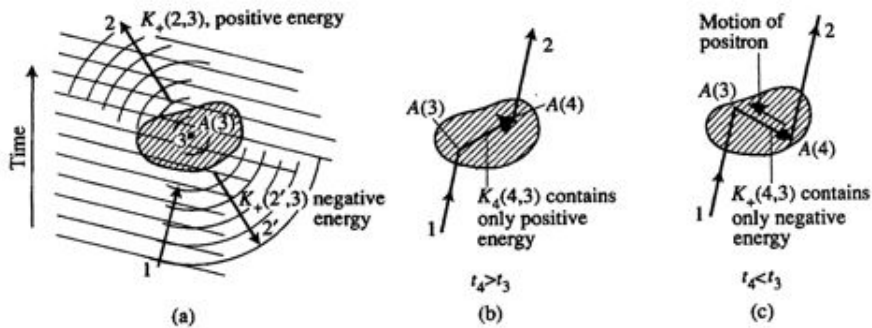

Figure 3. Different electron scattering possibilities from a potential region. (a) is a first order process while (b) and (c) are second order. [Based on Figure 2 of R. P. Feynman, "Theory of Positrons", *Phys. Rev.* **76**, 749-759 (1949)]

In a detector such as the bubble or cloud chambers of the past, where the path followed by the charged particles is made visible by repeated localizing interactions with the medium, one would observe a pair creation event at A(4), an electron coming in from the lower left, and an annihilation event at A(3). Of course, since the particles involved here are massive, in the case of real pair production the interval between A(3) and A(4) is time-like and the spatial distance between these events depends on the observer.

To reiterate, a world line is a classical concept that is only approximated in quantum mechanics by the kind of repeated interactions that make a path visible in a detector.[33] Minkowski space is the space of *events*—drawing a world line in a Minkowski diagram implicitly assumes such repeated interactions taken to the limit of the continuum.[34] While the characterization of Minkowski space as the space of events is often obscured by drawing world lines as representing the putative path of a particle in spacetime independent of its interactions, remembering that each point in Minkowski space is the position of a potential event removes much of the apparent incompatibility between quantum mechanics and special relativity, but it leaves us with a revised view of what constitutes motion.

**Quantum Mechanical Motion**

The picture of motion that emerges after the melding of quantum mechanics and special relativity is very unlike that of the classical picture of the path of a massive particle—like a marble—moving in spacetime. Consider a Minkowski diagram showing the world lines of several marbles at different locations. Given a space-like hypersurface corresponding to an instant of time in some frame, all the marbles would be visible at some set of locations. If one chooses a neighboring instant of time, these marbles would all still be visible at slightly different locations. This is because of the sharp localization of the marbles in space and time due to the continual interactions of their constituent components. Now consider the case of several elementary particles such as electrons. On any space-like hypersurface, the only particles "visible" would be those that were localized by an interaction to a region of spacetime that included the instant of time corresponding to the hypersurface.[35] After any localization, the wave function of a particle spreads both in space and in either direction in time. Consequently, neighboring hypersurfaces (in the same reference frame) corresponding to slightly different times could have a different set of particles that were "visible." If motion consists of a sequential series of localizations along a particle's path, it is not possible to define a continuum of movement in the classical sense—there exists only a series of "snapshots."

---

[33] Because the discussion to follow will give a different picture of a particle path, this is a good point to illustrate how motion is often described in quantum mechanics. Bohm [D. Bohm, *Quantum Theory*, Prentice-Hall, Inc., N.J. 1961, p.137.] in describing how a particle path is produced in a cloud chamber maintains that ". . . when the electron wave packet enters the chamber, it is quickly broken up into independent packets with no definite phase relation between them . . . the electron exists in only *one* of these packets, and the wave function represents only the *probability* that any given packet is the correct one. Each of these packets can then serve as a possible starting point for a new trajectory, but each of these starting points must be considered as a separate and distinct possibility, which, if realized, excludes all others." If the particle has large momentum, ". . . the uncertainty in momentum introduced as a result of the interaction with the atom results in only a small deflection, so that the noninterfering packets all travel with almost the same speed and direction as that of the incident particle." [emphasis in the original]

[34] There is a considerable—and quite interesting—literature dealing with repeated "measurements" of a particle and what is known as "Turing's Paradox" or the "Quantum Zeno Effect." See, for example: B. Misra and E. C. G. Sudarshan, *J. Math. Phys.* **18**, 756 (1977); D. Home and M. A. B. Whitaker, *Ann. of Phys.* **258**, 237 (1997), lanl.arXiv.org, quant-ph/0401164.

Haag,[36] has put this somewhat different terms: "The resulting ontological picture differs drastically from a classical one. It sketches a world, which is continuously evolving, where new facts are permanently emerging. Facts of the past determine only probabilities of future possibilities. While an individual event is considered as a real fact, the correlations between events due to quantum mechanical entanglement imply that an individual object can be regarded as real only insofar as it carries a causal link between two events. The object remains an element of potentiality as long as the target result has not become a completed fact."

It is important to emphasize that between localizations due to interactions, an elementary particle does not have a specifiable location, although—because it is located with very high probability[37] somewhere within the future and past light cones associated with its most recent localization—it would contribute to the local mass-energy density. This is not a matter of our ignorance, it is a fundamental property of quantum mechanics; Bell's theorem tells us that there are no hidden variables that could specify a particle's position between localizations.

---

[35] The term "visible" is put in quotes as a short-hand for the physical processes involved: the interaction of the particle needed to localize it on the space-like hypersurface and the detection of that interaction by the observer. It should also be emphasized that localization is in both space and *time*. Just as localization in space to dimensions comparable to the Compton wavelength corresponds to an uncertainty in momentum of ~$mc$, localization in time must be $\geq h/mc^2$ if the uncertainty in energy is to be less than or equal to the rest mass energy. For electrons this corresponds to $\geq 10^{-20}$ second.

[36] Rudolph Haag, Quantum Theory and the Division of the World, Mind and Matter **2**, 53 (2004).

[37] If one uses only positive energy solutions of the Dirac equation to form a wave packet, as noted earlier, the probability of finding a particle outside the light cone nowhere vanishes, although the propagator becomes very small for distances greater than the Compton wavelength $\hbar/mc$. See also the discussion in R.P. Feynman, *op .cit.* (1962).

As an example of how localization works, consider a single atom. Its nucleus is localized by the continuous interactions of its constituent components. The electrons are localized due to interactions with the nucleus, but only up to the appropriate quantum numbers—*n, l, m*, and *s*. One cannot localize the electrons to positions in their "orbits."

Earlier, we defined massive quantum mechanical particles as an excitation of spacetime. But in particle physics the concept of an elementary particle is intimately involved with group theory. In classical and quantum mechanics geometrical transformations—either Galilean or special relativistic—do not change what we consider to be the intrinsic properties of a particle. What this means of course is that there is a group property associated with the particle. The group of particular interest for quantum mechanics is the Poincaré group. The standard model of particle physics has enlarged this group, but the idea that a particle is associated with its group transformation properties—introduced by Wigner[38] over fifty years ago—remains unchanged.

There is also the issue of the transition from the domain where quantum mechanical descriptions are necessary to the classical world. While it is clear that the transition from quantum mechanics to classical mechanics corresponds, in some sense, to letting $\hbar$ go to zero, how the connection is made between Poisson brackets, which are concerned with coordinates in phase space, and commutators, which are operators on some Hilbert space, is muddy at best. The same is true for the transition from the canonical transformations of classical mechanics to the unitary transformations of quantum mechanics. An introductory discussion of the concept of an elementary particle and the transition between classical and quantum mechanics is given in Appendix II.

The advent of quantum mechanics mandated that the classical notion of an elementary particle be given up. In the end, we must live with the fact that elementary particles are some form of spacetime excitation that can be localized through interactions and even when not localized obey all the relevant conservation rules and retain "particle" properties such as mass, spin, and charge.

---

[38] E. P. Wigner, *Ann. Math.* **40**, 149 (1939).

Above, the flat spacetime of special relativity was used for the purpose of discussion. When the spacetime curvature due to gravitation is included, Minkowski diagrams become almost impossible to draw: Given a space-like hypersurface, the rate of clocks at any point on the hypersurface depends on the local mass-energy density and on local charge. Compared to a clock in empty spacetime, a clock near a concentration of mass-energy will run slower and will run faster near an electric charge of either sign.[39] Thus the hypersurface does not remain "planar" as it evolves in time. To draw world lines one must take into account the general relativistic metric. This is why one uses light cone indicators at points contained in regions of interest.

The concepts of quantum mechanical localization and the resulting picture of motion are especially important in discussing many-particle problems and the transition to the classical world.

The fundamental forces between the elementary particles are described today in terms of gauge fields. For example, the key concept for representing the electromagnetic force as a gauge field is the recognition that the phase of a particle's wave function must be treated as a new physical degree of freedom dependent on the particle's spacetime position. The four-dimensional vector potential plays the role of a connection relating the phase from point-to-point. Thus, the vector potential becomes the fundamental field for electromagnetism. Recall that the quantum mechanical wave function was called a phase wave by de Broglie.

A free particle at rest samples a volume of space *at least* as large as its Compton wavelength, and the wave function associated with this sampling is such that a spherical volume is sampled in the absence of external forces.  If a force acts on the particle, the probability distribution associated with its wave function is accordingly modified.  Say the force acts along the $x$-axis—the spherical symmetry is broken by an extension of the probability distribution (the volume sampled) along the $x$-axis.  As discussed above, to actually be "seen" to move, the particle must participate in a series of interactions so as to repeatedly localize it along its path of motion.  If the force acting on the particle is modeled as a virtual exchange of quanta, such an exchange—viewed as an interaction—would serve to localize the particle.  This concept for the action of forces must also be true for macroscopic objects, although now the description is far more complicated by the structure of matter and associated surface physics.[40]

---

[39] G.E. Marsh, "Charge, Geometry, and Effective Mass", *Found. Phys.* **38**, 293-300 (2008); "Charge, Geometry, and Effective Mass in the Kerr-Newman Solution to the Einstein Field Equations", *Found. Phys.* **38**, 959-968 (2008).

**Space, Time, and Matter**
Little is known about the empty spacetime continuum itself—or vacuum in the context of quantum field theory—except for what hints we have from special and general relativity, and those given by the Standard Model of particle physics.  Unfortunately, the greatest fundamental conceptual issue with the Standard Model is that its redefinition of the vacuum begins to make it look like some form of æther, albeit a relativistic one!  This results from the imposition of analogies from condensed matter physics, and in particular superconductivity.  Surely these analogies should not be taken literally.  The fact that they "work" should only be taken as a hint about the real nature of the vacuum.

When we say we "understanding" something we generally mean we can relate it to something simpler that we already understand; and in the case of spacetime, this usually means quantum mechanics.  And many attempts have been made to do this, none with outstanding success.  All are based on the idea that general relativity tells us that spacetime is a dynamical entity, while quantum mechanics tells us that a dynamical entity has quanta associated with it, and consequently this entity can be in a superposition of quantum states.  The implication is that there are "quanta" of space and time.  But what does this mean? Does it mean that space is made up of little elemental parcels of three-dimensional space?  What role would time play with such parcels?  Are there four-dimensional parcels of spacetime? Is time itself infinitely divisible? If not, is it made up of minimal steps? Is the ordering of such steps fixed?

---

[40] G.E. Marsh (2010), *op. cit.*

The usual approach to quantum gravity is to treat the dynamical variable as being the spacetime metric $g_{ij}(x)$. Then the usual procedure of quantization leads to the infamous Wheeler-DeWitt equation, which DeWitt was known to refer to as "that damned equation". The Wheeler-DeWitt equation is essentially the Schrödinger equation for the gravitational field, and its wavefunction, $\Psi[g_{ij}(x)]$, is the "wavefunction of the universe". Time does not explicitly appear in the equation and there are conceptual problems with regard to the definition of probability, not to speak of the fact that the resulting theory is not renormalizable.

An analogy that may help with regard to these questions is to represent spacetime as a piece of cloth: from a distance it is quite smooth, but as one comes closer it begins to show the structure of its weave. The argument is made that if we look at space and time at the Planck distance and time, it would show a structure that we could understand and use to explain the nature of spacetime. It is string theory and loop quantum gravity that attempt to address these questions.

With regard to "Matter" the Standard Model—despite questions about the nature of the vacuum alluded to above—has given us a great understanding of the relationships between its components. But there remain many questions that are not answered in the Standard Model: Why are there three families of quarks and leptons? What is the relationship, if any, between quarks and leptons? There are three arbitrary coupling constants associated with the constituent gauge groups of the Standard Model whose value has to be put in by hand. Because the Weinberg mixing angle is arbitrary, there is significant mixing—making the weak and electromagnetic forces appear related—only because experiment shows the coupling constants are of the same order of magnitude. The situation would be different if the mixing angle was close to zero or $\pi/2$. The quantization of charge is not explained since it is put into the theory arbitrarily when assigning values to weak hypercharge. The Standard Model requires only one Higgs boson, but going beyond the model there may be an expanded "Higgs sector" with a number of Higgs bosons, neutral as well as charged. At this point there is no *strong* evidence for an expanded Higgs sector. In the Standard Model, neutrino masses are zero; yet there is good experimental evidence for small neutrino masses and for neutrino oscillations—where neutrinos change their flavor. The most popular approach to these problems is to assume the fields of the Standard Model are fundamental, but that they are related by additional symmetries that are broken at higher energy scales. None have yet proved satisfactory.

Being able to explain the motion of matter through spacetime is one of the great successes of quantum mechanics, but Weyl's claim that "A definite portion of matter occupies a definite part of space at a definite moment of time," is a classical statement that is at best a limiting case of the quantum world. His claim that "It is in the composite idea of motion that these three fundamental conceptions [space, time, and matter] enter into intimate relationship," is certainly true, but this relationship does not offer much of an insight into the ultimate nature of space, time, or matter.

• • •

In the end it may be that there are limitations to the phenomenological approach of science to addressing epistemological or metaphysical issues. Readers should decide for themselves whether the situation with regard to our understanding of space and time has moved significantly beyond the following portion of the ~1959 lecture of Professor Walter von der Vogelweide:[41]

> Introduction: "And now, ladies and gentlemen, Professor Walter von der Vogelweide will present *A Short Talk On The Universe*.
>
> Now, why, you will ask me, have I chosen to speak on the Universe rather than some other topic. Well, it's very simple. There isn't anything else!

---

[41] From Severn Darden's *A Short Talk on the Universe*. This portion of professor von der Vogelweide's talk can be heard by clicking on http://www.gemarsh.com/wp-content/uploads/SpaceTimeM.mp3. The kind of improvisation that this slightly edited extract comes from began in the back of a bar near the University of Chicago campus called the *Compass*. The *Compass Players*, including Mike Nichols, Elaine May, Shelley Berman, and Severn Darden, performed in Hyde Park from 1955-1958 and several of the members went on to form *The Second City Theater* in 1959.

Now, in the universe we have time, space, motion, and thought. Now, you will ask me, what is this thing called time? [several second pause] *THAT* is time.

Now, you will ask me, what is space? Now this over here—this is some space. However, this is not all space. However, when I said that was time that was all the time there was anywhere in the universe—at that time. Now, if you were to take all of the space that there is in the universe and *CRAM* it into this little tiny place, this would be *ALL* the space there was! Unless of course, some leaked out. Which it could. And did! Hence the universe!"

# Appendix I: The Cosmic Microwave Background Radiation

The cosmic microwave background radiation is extremely important in cosmology because, in addition to the expansion of the universe seen from the red shift of light from distant galaxies, it shows that the universe had a very hot beginning. The radiation has an almost perfect black body spectrum of ~2.8 degrees *K* as seen below:

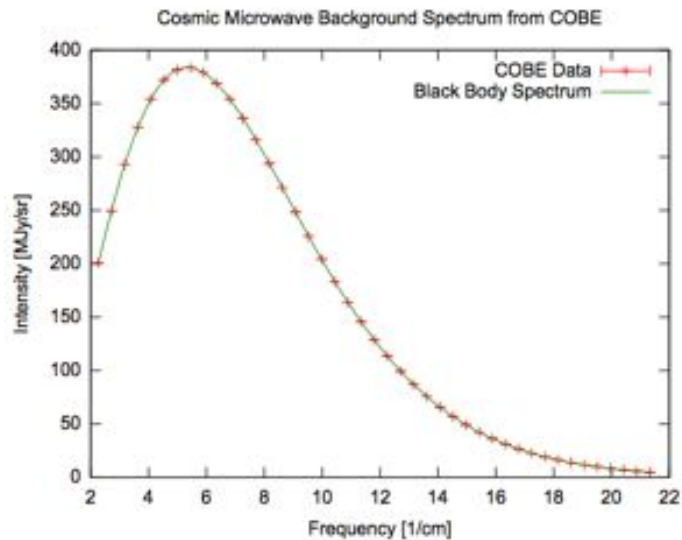

One often sees the very small *variations* in this background radiation plotted against a map of the sky known as the celestial sphere. It looks like the figure below:

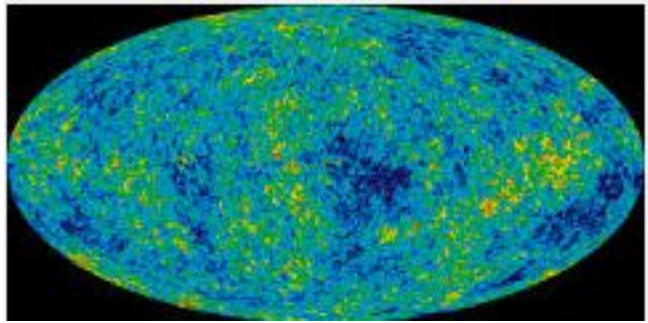

One can gain a great deal more information from the data contained in this map by plotting them as a spherical harmonic decomposition known as the power spectrum of the spatial fluctuations. This is shown in the figure below. The key to understanding this figure is understanding that the temperature variance associated with the $l^{th}$ multipole moment in the figure measures the mean-square temperature difference between points on the celestial sphere separated by angles of $180°/l$. Specifically, the temperature variance is given by $l(l+1)C_l/2\pi$, where $C_l$ is the coefficient of the $l^{th}$ multipole moment, which measures the mean-square temperature variance and has the units $\mu K^2$. Note that the temperature is squared.

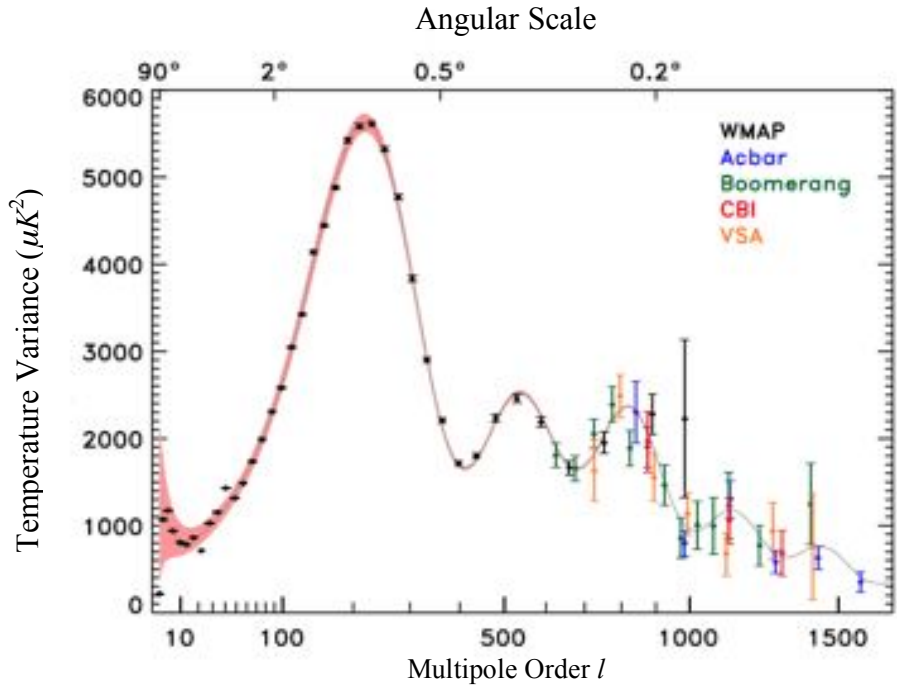

While it cannot be readily seen in this figure, there is a significant discrepancy between the observed temperature variance at the $l = 2$ quadrupole compared to the theoretical value. This is confirmed by the next few harmonics,[42] which are also not visible. This discrepancy is inconsistent with the scale-invariance predicted by some inflationary scenarios for the observed temperature fluctuations.

The peaks in this figure are due to "acoustic" compressional waves in the viscous-elastic "fluid"—primarily composed of hydrogen and helium ions, electrons, and photons—that characterized the plasma epoch of the universe around 400,000 years after the "Big Bang". The "acoustic" oscillations are also known as gravitationally driven photon-baryon oscillations.

---

[42] R. Penrose, *The Road to Reality* (Alfred A. Knopf, New York 2005), §28.10.

The first peak is at an *l* of about 200 corresponding to an angular scale of about $1^o$ on the sky; that is, $180^o/l \sim 1^o$. This is what would be expected if the path of the greater than 10 Gyr journey of the radiation from ~400,000 years after the "Big Bang" were not distorted by the geometry of the universe; that is, the *position* of the first and largest peak tells us that the universe is flat—it has no curvature.

The height of the second peak (really a cold spot peak, as are the $4^{th}$, $6^{th}$, etc. peaks—remember, the temperature is squared in the plot) relative to first peak turns out to be a sensitive measure of the baryon density in the plasma.[43]

The study of type Ia supernovae, thought to represent a "standard candle" throughout the universe, tells us that the expansion of the universe has actually been speeding up for the last ~5 Gyr. This implies that the vacuum itself has an energy density associated with it. If this is designated as $\Omega_\Lambda$; the total mass density of the universe as $\Omega_M$; the normal matter baryon density as $\Omega_B$; and we normalized the total to that for a flat universe ($\Omega_{TOT} = 1$), as required by the first peak in the power spectrum of the spatial fluctuations of the cosmic background radiation, then the composition of the universe would be given by $\Omega_{TOT} = \Omega_\Lambda + \Omega_M + \Omega_B = 1$.

The breakdown, as currently understood, is as follows: $\Omega_M = 0.24$ of which normal matter $\Omega_B = 0.04$ makes up only a small part, and $\Omega_\Lambda = 0.76$. So the universe is made up of about ~4% normal matter, ~20% "dark" matter, whose composition is unknown but has so far been found to interact only through gravitation, and ~76% of so called "dark energy", indistinguishable from Einstein's cosmological constant.

---

[43] Teasing out the baryon density from the data is non-trivial, and there are many publications on the subject. One nice discussion is given in the thesis by Reijo Keskitalo available at: https://helda.helsinki.fi/bitstream/handle/10138/20988/theeffec.pdf?sequence=4

# Appendix II: Elementary Particles
# and
# The Quantum ⇔ Classical Transition

Wigner showed that a particle was to be associated with its transformation properties under the Poincaré group. In particular, the physically relevant representations of the Poincaré group for energies greater than or equal to zero are parameterized by

$$m^2 > 0 \qquad s = 0, \tfrac{1}{2}, 1, \tfrac{3}{2}, \ldots$$
$$m^2 = 0 \qquad s = 0, \pm\tfrac{1}{2}, \pm 1, \pm\tfrac{3}{2}, \ldots$$

where $m$ is the mass and $s$ the spin. Thus, each kind of elementary particle is associated with a unitary irreducible representation of the Poincaré group. In a real sense, the particle and the representation are identified. As put by Sternberg,[44] "an elementary particle 'is' an irreducible unitary representation of the group, $G$ of physics, where these representations are required to satisfy certain physically reasonable restrictions . . . ."

While the invariance of the intrinsic properties of a particle under the Poincaré group applies equally well in classical and quantum mechanics, irreducible representations are usually only associated with a particle in quantum mechanics since spin is not quantized in classical mechanics. But, as pointed out by Bacry,[45] Wigner did not restrict his approach to elementary particles, but referred to *elementary systems*. The example of an elementary system given by Bacry is that of the spin zero hydrogen atom in its ground state with mass somewhat less than the sum of the proton and electron masses. While the set of all states of the hydrogen atom form a representation space for a reducible representation of the Poincaré group, the proton and electron comprising the system no longer have irreducible representations associated with them since these particles are interacting and therefore do not form an isolated system.

---

[44] S. Sternberg, *Group Theory and Physics* (Cambridge University Press 1994), Sect. 3.9. Sternberg calls this Wigner's central "dogma".

[45] H. Bacry, *Localizability and Space in Quantum Physics*, Lecture Notes in Physics, No. 308 (Springer-Verlag, Berlin 1988), Ch. 3; *Commun. Math. Phys.* **5**, 97 (1967).

The lesson to be learned from the above example is that collections of elementary particles in a particular state, while they may continue to be associated with an irreducible representation of the appropriate group, may lose some group properties like spin that are purely quantum mechanical in nature. What remains is the mass of the aggregate system. Going in the direction of decreasing mass, Haag[46] has pointed out that "The physical interpretation of an irreducible representation of the Poincaré group (Newton and Wigner 1949) shows that the notion of a localized state of a particle becomes increasingly blurred with decreasing rest mass." Put the other way around, the localization of a particle is increasingly sharp as the mass increases. This can also be seen from the form of the Newton-Wigner position operators.

Haag's observation clearly relates to the transition from the quantum world to the classical one. It is generally thought that the relation between classical and quantum mechanics is characterized by "letting $\hbar$ go to zero." For example, a standard problem in textbooks is to show that

$$\lim_{\hbar \to 0} \frac{1}{i\hbar}[A,B] = \{A,B\}, \tag{1}$$

where [A, B] is the commutator and {A, B} the Poisson bracket. This relationship applies independent of mass. Note that the left hand side of this equation is an operator on a Hilbert space and the right hand side is a function.[47] It holds for most operators provided the Poisson bracket is considered to be an operator. And while there are some caveats,[48] it always holds in the classical limit.

Thus, there are at least two ways to go to the classical limit: the first is to let $\hbar \to 0$ as in Eq. (1); and the second is to go to the limit of large mass. For large mass, this equation reduces to $[A, B] = 0$ except for the case where $A = q$ and $B = p$, in which case one gets $[q, p] = i\hbar$ since $\{q, p\} = 1$.

---

[46] Rudolph Haag, *Quantum Theory and the Division of the World*, Mind and Matter **2**, 53 (2004); T. D. Newton and E. P. Wigner, *Rev. Mod. Phys*. **21**, 400 (1949).
[47] The momentum and position take the form of operators on the l.h.s. of this equation and coordinates in phase space on the r.h.s.
[48] See, for example, D. Bohm, *Quantum Theory* (Prentice-Hall, Inc., N.J. 1951), Sect. 16.23.

The original derivation of this equation was given by Dirac,[49] but before turning to Dirac's derivation of this equation, consider the non-commuting matrices $U$, $V$, $U_1$, $U_2$, $V_1$, $V_2$. It is readily shown that the commutators $[U,V_1V_2]$ and $[U_1U_2,V]$ are

$$[U,V_1V_2] = [U,V_1]V_2 + V_1[U,V_2]$$

$$[U_1U_2,V] = U_1[U_2,V] + [U_1,V]U_2. \qquad (2)$$

Thus, the commutators on the left hand side of these equations automatically satisfy the Leibniz rule. Dirac, in his derivation of Eq. (1) begins with Poisson brackets and when he arrives at the analog of Eqs. (2), holds the order of the corresponding commuting dynamical variables fixed; i.e., having satisfied the Leibniz rule, he henceforth treats these variables *as if they were non-commuting matrices*.

To be quite explicit, Dirac obtains the equations

$$\{u,v_1v_2\} = \{u,v_1\}v_2 + v_1\{u,v_2\}$$
$$\{u_1u_2,v\} = u_1\{u_2,v\} + \{u_1,v\}u_2, \qquad (3)$$

and then *requires that the order of $u_1$ and $u_2$ be preserved in the second equation and the order of $v_1$ and $v_2$ in the first*. Dirac now evaluates $\{u_1u_2, v_1v_2\}$ in two ways using Eqs. (3), and subsequently equates the result to obtain

$$\{u_1,v_1\}[u_2,v_2] = [u_1,v_1]\{u_2,v_2\}. \qquad (4)$$

Since $u_1$ and $u_2$ are independent of $v_1$ and $v_2$, Eq. (4) implies that

---

$$[u,v] = i\hbar\{u,v\}. \tag{5}$$

The value of the constant $\hbar$ is set by experiment and the factor $i$ is introduced for the following reason: Dirac treats $u$ and $v$ as linear operators that could have an imaginary part and since the product of two real (i.e., Hermitian) operators is not necessarily real—unless they commute, Dirac introduces the factor of $i$ to guarantee that $i(uv - vu)$ is real.

Instead of using Dirac's mixed approach of arbitrarily fixing the order of $u_1$ and $u_2$ and $v_1$ and $v_2$ as above, one can begin by initially treating these variables as non-commuting matrices in the Poisson bracket—some matrix representation of the invariance group. Treating $U, V, U_1, U_2, V_1, V_2$ as matrices results in

$$\{U,V_1V_2\} = \{U,V_1\}V_2 + V_1\{U,V_2\} \quad provided \quad \left[\frac{\partial U}{\partial q},V_1\right] = \left[\frac{\partial U}{\partial p},V_2\right] = 0$$

$$\{U_1U_2,V\} = U_1\{U_2,V\} + \{U_1,V\}U_2 \quad provided \quad \left[\frac{\partial V}{\partial q},U_1\right] = \left[\frac{\partial V}{\partial p},U_2\right] = 0. \tag{6}$$

These correspond to Dirac's equations given by Eqs. (3). Note that the vanishing of the commutators on the right hand side of Eqs. (6) guarantees that the Poisson brackets on the left side obey the Leibniz rule. If $\{U_1U_2,V_1V_2\}$ is now evaluated *à la* Dirac, Eq. (5) is again obtained.

Thus, the requirements imposed by Dirac to derive Eq. (5) are equivalent to starting with non-commuting variables in the Poisson bracket to find a set of commutators whose vanishing guarantees that the Poisson brackets obey the Leibnitz rule.

There is another approach to showing the relationship between classical and quantum mechanics and that is to introduce complex canonical coordinates. This path is attractive because it seemingly allows both classical and quantum mechanics to be formally embedded in the same mathematical structure.

Consider first Hamilton's equations.

$$\dot{q}_j = \frac{\partial H}{\partial p_j}, \quad \dot{p}_j = -\frac{\partial H}{\partial q_j}. \tag{7}$$

Now define the following complex coordinates and partial derivatives:

$$z_j = \frac{1}{2}(q_j + ip_j), \quad z_j^* = \frac{1}{2}(q_j - ip_j)$$
$$\frac{\partial}{\partial z} = \frac{1}{2}\left(\frac{\partial}{\partial q_j} - i\frac{\partial}{\partial p_j}\right), \quad \frac{\partial}{\partial z^*} = \frac{1}{2}\left(\frac{\partial}{\partial q_j} + i\frac{\partial}{\partial p_j}\right). \tag{8}$$

A little algebra then shows that Eqs. (7) corresponds to

$$\dot{z} = -i\frac{\partial H}{\partial z^*}. \tag{9}$$

We may obtain formally the same expression in quantum mechanics: The Schrödinger equation is

$$H|\Psi\rangle = i\hbar \partial_t |\Psi\rangle. \tag{10}$$

If we now expand $|\Psi\rangle$ in terms of the fixed basis kets $|\alpha_k\rangle$ with complex coefficients $u_k$, we obtain

$$\langle \alpha_i | H | \alpha_k \rangle u_i^* u_k = i\hbar \delta_{ik} u_i^* \partial_t u_k. \tag{11}$$

It is the change in these coefficients that correspond to the evolution of the system in time. Now if we define $\mathcal{H} = \langle \alpha_i | H | \alpha_k \rangle u_i^* u_k = \langle \Psi | H | \Psi \rangle$ and set $k = i$, Eq. (11) becomes

$$\mathcal{H} = i\hbar u_i^* \partial_t u_i. \tag{12}$$

Thus,

$$\hbar \dot{u}_i = -i\frac{\partial \mathcal{H}}{\partial u_i^*}. \tag{13}$$

If we choose units where $\hbar = c = 1$, Eq. (13) is formally the same as Eq. (9).

Strocchi[50] introduced complex coordinates in order to address the fact that "The classical limit of quantum mechanics, which is usually identified with the limit $\hbar \to 0$, is rather obscure; the connection between commutators and Poisson brackets is difficult to explain in that limit. Neither is the connection between the theory of canonical transformations and unitary transformations in quantum mechanics apparent, and one has to rely on analogy arguments." The paper demonstrates that $\{\mathcal{A},\mathcal{B}\} = \langle[A,B]\rangle$; i.e., that the classical Poisson bracket between the quantities $\mathcal{A}$ and $\mathcal{B}$, what Strocchi calls classical phase functions, is the mean value of the commutator between the corresponding operators; it was also shown that unitary transformations in quantum mechanics are canonical transformations in the sense of classical mechanics.

---

[50] F. Strocchi, "Complex Coordinates and Quantum Mechanics", *Rev. Mod. Phys.* **38**, (1966).

The formalism using complex canonical coordinates is interesting because it shows that both classical and quantum mechanics can be embedded into a unified structure. The Schrödinger equation, as shown above, appears as Hamilton's equations in complex canonical coordinates for a classical system with the mean value of the quantum Hamiltonian operator being the Hamiltonian function.

The real question is whether the introduction of complex coordinates will lead to new results or insights or is simply an elegant reformulation of the problem of the classical to quantum transition.

# Appendix III: CP Violation and Baryogenesis[51]

The neutral *K*-meson, designated $K^0$, is composed of a down and an anti-strange quark and its antiparticle, the $\bar{K}^0$, is composed of an anti-down quark and a strange quark. Because there are no conserved quantum numbers—when weak interactions are taken into account—that distinguish the $K^0$ and $\bar{K}^0$, transitions (called mixing) are allowed between them. This is unusual and is not the case for charged particles and their antiparticles; e.g., positron-electron pairs or proton-antiproton pairs. There are two combinations, called the *K*-long, $K_L$, and *K*-short, $K_S$, after their decay times, that are observed and measured and should therefore be thought of as the "physical" *K*-mesons. These mixed states are *almost CP*-eigenstates that can be thought of as "oscillating" between the original $K^0$ and $\bar{K}^0$. A similar state of affairs occurs for the neutral *B*-meson and *D*-meson. The $K_L$ and $K_S$ states are not antiparticles of each other and are therefore not required to have the same mass or lifetimes as is required by relativistic quantum mechanics for particle-antiparticle pairs. Because the $K_L$ and $K_S$ states consist of approximately an equal mixture of the $K^0$ and $\bar{K}^0$ states, the *CP* violation of the neutral *K*-meson is very small. This small deviation is enough, however, to result in a small difference in their mass, which determines the frequency of oscillation between the $K^0$ and $\bar{K}^0$ states.

In the standard model of particle physics, *CP*-violation has its origin in quark mixing and is characterized in what is called the CKM matrix, some of whose elements contain a phase angle, which if it is nonzero, implies *T*-violation and *CP*-violation if the *CPT* theorem is assumed. The *CP*-violation described here cannot be the only source of violation in the early universe because the matter-antimatter asymmetry observed today requires *CP*-violation of several orders of magnitude greater than that needed to account for meson decays. There have been numerous proposals to explain baryogeneses including "spontaneous baryogenesis", which does not require thermodynamic non-equilibrium conditions (see the book by Kolb and Turner), and there is even the possibility of introducing torsion into the spacetime metric.[52]

---

[51] There are a number of excellent texts that are relevant to this section. Among them are: B.R. Martin and G. Shaw, *Particle Physics* (Wiley and Sons Ltd, 2008); G. Castelo Branco, L. Lavoura, and J. P. Silva, *CP Violation* (ClarendonPress, Oxford 1999); E.W. Kolb and M.S. Turner, *op. cit.*

The empty spacetime continuum is known as the vacuum in the context of quantum field theory. What we know about this vacuum comes from the hints we have from special and general relativity, and those given to us by the Standard Model of particle physics. Since the particles that make up the matter in the universe are created from this vacuum, and some of these particles violate time invariance, it would make sense to consider the possibility that violations of time reversal symmetry reflect a basic property of the vacuum and that time asymmetry is an intrinsic property of empty spacetime itself. Another source of time asymmetry in the very early universe could be asymmetries in initial conditions or more general cosmological boundary conditions.

In quantum field theory, the vacuum is invariant under the discrete symmetries of *C*, *P*, and *T*. One might postulate that these symmetries are broken in the very early universe to a greater extent than we see in the violation of *CP*-invariance in particle physics. This would occur above a critical temperature $T_C$, during the period when one might expect quantum gravity to apply. This approach to the problem goes under the name of baryogenesis via spontaneous Lorentz violation.[53] The very small breaking of *CP*-invariance, and hence the breaking of time reversal invariance responsible for the asymmetry in time, seen in the universe when the temperature falls below $T_C$ might be a remnant of this earlier period.

The value of the critical temperature $T_C$ sets the time that this kind of spontaneous symmetry breaking occurred as well as the temperature. The Planck time, determined by the dimensions of the fundamental constants of nature, is about $10^{-44}$ seconds at which time the temperature was $10^{32}$ °K and the diameter of the universe $10^{-3}$ cm. $T_C$ would presumably lie between this temperature and that of the electroweak phase transition at $10^{-10}$ seconds when the temperature was $10^{15}$ °K and the diameter of the universe $10^{14}$ cm.

---

[52] N.K. Poplawski, Neutral-meson oscillations with torsion, arXiv: 1105.0102v1 [hep-ph] 30 April 2011.
[53] See, for example: S.M. Carroll and J. Shu, "Models of Baryogenesis via Spontaneous Lorentz Violation", arXiv: hep-ph/0510081v3 30 Nov 2005.

A similar proposal has been put forth by Moffat[54] who postulates that, "The local Lorentz and diffeomorphism symmetries of Einstein's gravitational theory are spontaneously broken by a Higgs mechanism by invoking a phase transition in the early Universe, at a critical temperature $T_c$ below which the symmetry is restored. . . . The time direction of the vacuum expectation value of the scalar Higgs field generates a time asymmetry, which defines the cosmological arrow of time and the direction of increasing entropy as the Lorentz symmetry is restored at low temperatures. . . . The presence of the Lorentz symmetry broken phase at high temperatures will spontaneously create matter at the beginning of the Universe, due to the violation of the energy conservation. This could explain the origin of matter in the early Universe." Note that Moffat conflates here the cosmological arrow of time and the direction of increasing entropy.

There is one niggling point that must be addressed: As I noted above, in quantum field theory, the vacuum is invariant under the discrete symmetries of $C$, $P$, and $T$. But time in our three-dimensional space has an asymmetry that results from the expansion of the universe. Moreover, the time coordinate used to express physical relationships is identified with this cosmic time coordinate. The question then arises as to whether this asymmetry is related to the occurrence of $CP$-violation,[55] and therefore time reversal invariance violation, in the weak interactions as exemplified by the decay of the $K^0$ or $B$ mesons. While the $T$-operator has nothing to do with thermodynamic time, it is not obvious that there is no relation between it and cosmic time. The $T$-operator transforms state vectors and in quantum mechanics such transformations are usually restricted to being unitary. If one considers the Schrödinger equation, the transformed state vector will only satisfy this equation if the Hamiltonian is invariant and the *T*-operator is antiunitary. Under such transformations, the time variable $t$ in the equation is transformed to $-t$. But this time variable is that of cosmic time. Because of these relationships, one should at least consider the possibility that the time reversal invariance violation of the $K^0$ or *B* mesons has its origin in the asymmetry of cosmic time.

---

[54] J. W. Moffat, "Quantum Gravity, the Origin of Time and Time's Arrow", *Foundations of Physics*, **23**, 411 (1992). gr-qc/9209001. In this connection see also George F R Ellis, "The arrow of time and the nature of spacetime", arXiv:1302.7291v2 (2013).
[55] An extensive discussion of the topic of $CP$ and $T$-violation is given in: R. G. Sachs, *The Physics of Time Reversal* (University of Chicago Press, Chicago 1987).